\documentclass{article}
\pdfoutput=1
\usepackage[T1]{fontenc}
\usepackage[utf8]{inputenc}
\usepackage[square,sort,comma,numbers]{natbib}
\usepackage{authblk}
\usepackage{color}
\usepackage{graphicx}
\usepackage{epstopdf}
\usepackage{algorithmic}
\usepackage{algorithm}
\usepackage{paralist} 
\usepackage{tabularx}
\usepackage{fancyhdr}

\newcommand{\lqu}{~\lq\lq}
\newcommand{\rqu}{\rq\rq}

\title{A stochastic model of catalytic reaction networks in protocells}
\author[1,2]{Roberto Serra}
\author[2]{Alessandro Filisetti}
\author[1,2]{Marco Villani}
\author[3]{Alex Graudenzi}
\author[3,4]{Chiara Damiani}
\author[5]{Tommaso Panini}
\affil[1]{Dept. of Physics, Informatics and Mathematics, Modena and Reggio Emilia University}
\affil[2]{European Centre for Living Technology, ECLT, University Ca' Foscari of Venice, Italy}
\affil[3]{Dept. of Informatics, Systems and Communication, University of Milan Bicocca, Italy}
\affil[4]{SYSBIO - Centre for Systems Biology, Piazza della Scienza 2, 20126 Milano, Italy}
\affil[5]{University of Turin, Department of Economics and Statistics "Cognetti de Martiis", Torino}

\begin{document}
\fancyhead[R]{R. Serra et al.}
\fancyhead[L]{\thepage}
\maketitle

\begin{abstract} Protocells are supposed to have played a key role in the self-organizing processes leading to the emergence of life. Existing models either $(i)$ describe protocell architecture and dynamics, given the existence of sets of collectively self-replicating molecules for granted, or $(ii)$ describe the emergence of the aforementioned sets from an ensemble of random molecules in a simple experimental setting (e.g. a closed system or a steady-state flow reactor) that does not properly describe a protocell. In this paper we present a model that goes beyond these limitations by describing the dynamics of sets of replicating molecules within a lipid vesicle. We adopt the simplest possible protocell architecture, by considering a semi-permeable membrane that selects the molecular types that are allowed to enter or exit the protocell and by assuming that the reactions take place in the aqueous phase in the internal compartment. As a first approximation, we ignore the protocell growth and division dynamics. The behavior of catalytic reaction networks is then simulated by means of a stochastic model that accounts for the creation and the extinction of species and reactions. While this is not yet an exhaustive protocell model, it already provides clues regarding some processes that are relevant for understanding the conditions that can enable a population of protocells to undergo evolution and selection. 

\end{abstract}


\section{Introduction}
\label{sec:intro}
It is widely believed that the origin of life required the formation of sets of molecules able to collectively self-replicate, as well as of compartments able to undergo fission and proliferate, i.e.{\em protocells}~\citep{Szostak:2001xw,Rasmussen:2004aa,Rasmussen2003}. In particular, in order to observe a lifelike behavior it  was necessary that some of the chemical reactions were coupled to the rate of proliferation of the compartments.

Several protocell architectures have been proposed~\citep{Carletti:2008rm,Filisetti2010b,Ganti:2003aa,Luisi2006,Mansy:2008kx,Morowitz1988,Munteanu:2006vn,Rasmussen2003,Rocheleau:2007gf,Segre2000,Serra:2007aa,Sole:2007jw,Stano2010,Szostak:2001xw}, most of them identifying the compartment with a lipid vesicle that may spontaneously fission under suitable circumstances. 

On the other hand, many distinct models were proposed to describe sets of reactions involving randomly generated molecules~\citep{R.J.Bagley:1991ys,Farmer1986,Hordijk2010,Kauffman:1986mi}. In many cases, although this is not in principle required, it is assumed that only {\em catalyzed} reactions take place at a significant rate, therefore these sets are also termed {\em catalytic reaction sets} (briefly, CRSs). It is worth noting that the appearance of new molecules implies the appearance of new reactions involving those new molecules, so that both the set of molecular types and the set of reactions change in time. Hence, it is possible that at a certain time a set of molecules able to catalyze each other's formation emerges~\citep{Farmer1986,Hordijk2010,Kauffman:1986mi}, and we will refer to it as an {\em autocatalytic set} (ACS). It can be noticed that a CRS can contain one or more ACSs, or none. \\

Even though some models of protocell actually describe the coupling between reaction networks and the dynamics of a lipid container, they consider only a fixed set of molecular species and reactions~\citep{Carletti:2008rm,Mansy2009,Serra:2006aa,Zhu:2009kx}, hence providing an incomplete representation of this complex interplay. Conversely, while there are several studies on collectively self replicating sets of molecules in a {\em continuously stirred open-flow tank reactor}, CSTR~\citep{R.J.Bagley:1991ys,hordijk-2002,Jain:2001xw,Stadler1991,Stadler1990,Stadler1995} including our own~\citep{Filisetti2011a,Serra2013,Filisetti2011b}, they provide only limited information about the behavior of a protocell.

Therefore, in order to develop a framework that may unify the CRSs and the protocell modeling approaches, it is necessary i) to analyze the behavior of CRSs in a vesicle, and ii) to investigate the coupling of the evolving chemical population with the growth of the lipid container and its fission. In this paper we propose a step towards the first goal, while deferring the second one to a further work.
In particular, we here analyze the behavior of a dynamical model of CRSs in a simplified model of a non-growing vesicle. To the best of our knowledge this is a novel approach.\\


A few important remarks. Let us first observe that the CSTR is not an {\em a-priori} good model of a protocell for at least two reasons: $i)$ in general, in protocells there is no constant inflow and $ii)$ protocells have semipermeable membranes, which allow the inflow/outflow of some, but not all, molecular types. On the contrary, in open flow reactors all that is contained in the inflow enters the reactor and all that is dissolved in the reactor can be washed out in the outflow.\footnote{Limitations on the outflow can be modeled in a chemostat e.g. by supposing that all the molecules that are larger than a certain size precipitate and cannot be washed away.}
Another important limit of the CSTR concerns its evolvability. It has been argued~\citep{Wesson1991,Morris2003} that the presence of different asymptotic dynamical states and the ability to shift between them may be essential to achieve the viable evolution of the first forms of life. Recent works~\citep{Vasas2012} have found that, in models of catalytic reaction networks in CSTRs, generally only one of these states is found, apart from fluctuations~\citep{Dai2009}.\\ 

Furthermore, in order to accomplish the goal of this work, we need to better specify both the model of catalytic reactions sets and that of the protocell.
As far as the former is concerned, we have studied the dynamics of random sets of molecules by revisiting a model by Kauffman, who proposed an interesting way to build new molecular species from the existing ones (see section~\ref{sec:model} for a description). The original version of the model relied on purely graph-theoretical arguments~\citep{Kauffman:1986mi}, which are important, but fail to appreciate the effects of the dynamics, including noise, fluctuations and small-number effects.
The dynamics has been later introduced by Farmer et al.~\citep{R.J.Bagley:1991ys,Farmer1986}, who described the kinetics by using ordinary differential equations. However, this formalism does not account for the chance of a species to become extinct in a finite amount of time, as it may instead well happen (so the reaction graph may grow but never shrinks). In order to overcome these limitations, Bagley proposed an empirical correction by setting to zero the concentration values that happen to fall below a certain threshold~\citep{R.J.Bagley:1991ys}. In our works we rather use from the very beginning a stochastic approach to analyze the dynamics, the well-known Gillespie algorithm~\citep{Gillespie:2007vn,Gillespie:1977fv}, in order to deal in a rigorous way with low concentrations and with their fluctuations.\\

Note that the Kauffman model largely relies upon randomness. In particular, every polymer in the system has a fixed probability (that may vanish) to catalyze any possible reaction. Therefore, in different simulations the same species can catalyze different reactions leading to the formation of different {\em\lqu  chemistries\rqu}. Thus, this is exactly the language we choose: a set of tuples $\left \{species, catalysis, reaction\right \}$, where the species catalyzes the reaction, will be called~\lqu chemistry\rqu, because it describes a possible artificial world.\footnote{It is worthwhile to notice that the presence of the\lqu catalysis\rqu~within the tuple allows the possibility for a species to catalyze more than one reaction and for a reaction to be catalyzed by more than one species.} We can then simulate different chemistries and look for generic properties of the set of chemistries; but in a different series of experiments we can also keep the chemistry fixed, and simulate various time histories. In principle, these may differ, since the discovery of a given catalyst at an early phase in a finite system might channel the following evolution in a way or another. Since the number of molecules of some species may be very small, it is not in principle legitimate to ignore this aspect, and our stochastic model is particularly well suited to analyze it, as it will be shown in Section~\ref{sec:model}. Of course, there can be conditions where all the simulations of a given chemistry converge asymptotically to the same chemical mixture.\\ 

Moving now to the protocell model, note that they are usually based on lipid vesicles, i.e. approximately spherical structures with an aqueous interior and a membrane composed by a lipid bilayer, which spontaneously form when lipids are mixed with water under certain conditions~\citep{Cans:2008ek,Dominak:2007gd,Hanczyc2003,Luisi2006,citeulike:1586749,Munteanu:2006vn,Polozova2005117}.
Even though different protocell architectures have been proposed, we will here consider the simplest model, namely that in which all the key reactions take place in the aqueous phase \textit{inside} the protocell. It would be indeed straightforward  to model the coupling between some of these molecules and the growth of the protocell following an approach similar to that of our previous studies~\citep{Carletti:2008rm,Filisetti2010b,Serra:2006aa}. Yet, the main objective of the present work is that of studying the dynamics of CRSs embedded in a vesicle, so we will simplify our treatment by ignoring the growth dynamics of the protocell, and keeping its volume fixed. This implies that our study will be limited to time intervals that are short with respect to those describing the growth of the whole protocell.\\

The selective character of membranes is a key ingredient of our model: we will suppose for simplicity that all (and only) the molecules that are shorter than a certain length can cross the membrane. The transmembrane motion of the permeable species is here supposed to be driven by the difference of their concentrations in the internal aqueous volume of the protocell and in the external aqueous environment. We will assume that transmembrane diffusion is extremely fast, so that there is always equilibrium between the concentrations of the species that can cross the membrane; this adiabatic hypothesis could be easily relaxed in the future.
Furthermore, we assume that protocells are turgid, so that the constant-volume approximation implies that we will also neglect issues related to osmotic pressure. Another related aspect of the model is that, since it is assumed that the permeable species are at equilibrium, while the non-permeable ones never cross the barriers, infinite concentration growth is possible; this is obviously a nonphysical behavior, so the model validity is limited in time. All these simplifications, which will be removed in subsequent studies, are also justified by the fact that our main goal is that of studying how the dynamics of CRSs are affected by being embedded in a vesicle.\\

This model can be used in order to investigate the behavior of the system in different conditions and to address some important questions. 
The first and perhaps most important one is the reason why compartments seem to be necessary for life. Indeed, the very first studies on self-replicating molecules~\citep{Eigen1977a,Dyson:1985uq,Kauffman:1986mi,Jain:1998fk} were not interested in this aspect, so the CRSs were supposed to exist, e.g., in a pond or in a beaker. Yet life seems to require compartments, that are ubiquitous.  It is then important to understand whether there are major differences between what may happen in a protocell and what happens in the bulk phase.
It would be unconvincing to postulate {\em a priori} that the internal and external environments are different. It is indeed more likely to assume that the vesicles form in a pre-existing aqueous environment, so the average internal milieu is essentially the same as the external one.\footnote{It has been observed that some superconcentration phenomena can take place under particular circumstances~\citep{DeSouza2012,Serra2013a} but we will neglect them here.} Then, if a membrane surrounds a portion of the fluid, what can happen that makes a difference?\\

Let us first observe that protocells are small (their typical linear dimensions ranging from $0.1$ to $10 \mu m$). If we imagine that a population of protocells exists, and they are not\lqu overcrowed\rqu, their total internal volume will typically be much smaller than the total external volume (this is {\em a fortiori} true for an isolated one). Moreover, every point in the interior of a protocell is not allowed to be far away from the surface of the protocell that contains it. These observations imply that the effect of surfaces will be much larger within protocells then outside them. Suppose for example that the membrane hosts some catalytic activities, so that important molecules are synthesized close to its boundaries, both inside and outside, and diffuse freely. If the membrane width is much smaller than the protocell radius, then the internal and external surface areas are very close to each other, but the external volume is much larger than the external one: therefore the internal concentrations will be much higher than those in the external environment. In this case, the system behavior in the interior can be significantly different from the external one~\citep{Serra2013a}. Note also that this effect may be different for different molecules: the formation of some of them might be catalyzed by the membrane, while others might be unaffected: so even the relative concentrations of different chemicals may differ in the two cases.\\

Indeed, there are important protocell models that are based on such an active catalytic role of the membrane~\citep{Rasmussen2003}. In these cases it is easy to understand what the role of the protocell is, since it provides essential catalysts and a way to keep their products closer. But protocells might be able to give rise to an internal environment different from the bulk even if the catalytic activity is absent.
The reason for this seemingly counterintuitive behavior is, once again, the smallness of the protocells. Note that we are considering a case where new molecules are formed (from those that are already in the interior of the protocell plus those that can cross the semipermeable membrane). If the concentrations are not too high, it is likely that the total numbers of newly formed molecules are quite low, so that different protocells might host different groups of molecules. It might even happen that a molecular type is present in some protocells and not in others.\\ 

In order to get a feeling for this possibility, let us provide some realistic estimates of the number of molecules of different types that can be present in a protocell.
Let us consider typical vesicles (linear dimension about $1\mu m$) and small ones ($0.1\mu m$). Typical concentrations of macromolecules may be in the millimolar to nanomolar range; the excepted numbers of molecules in a single protocell are therefore given in table~\ref{tab:expNumofMols}.
\begin{center}
\begin{table}
  \begin{tabular}{|c|c|c|c|c|c|}
    \hline
     & $1M$ & $1mM$ & $0.1mM$ & $1\mu M$ & $1nM$\\
    \hline
    Typical ($1 \mu^{3}$) & $10^{8}$ & $10^{5}$ & $10^{4}$ & $10^{2}$ & $0.1$\\
    \hline
    Small ($10^{-3} \mu^{3}$) & $10^{5}$ & $10^{2}$ & $10$ & $0.1$ & $10^{-4}$\\
    \hline
\end{tabular}
 \caption{excepted number of molecules of a given species in a given protocell; rows refer to protocell volumes, columns to concentrations.}
   \label{tab:expNumofMols}
 \end{table} 
  \end{center}
Let us recall that these numbers refer to the excepted values, but there are fluctuations that may be relevant when small numbers are involved. For example, in the case of a $1\mu M$ concentration in small vesicles, there will be on average $1$ molecule every $10$ cells: it is apparent that different protocells will have widely different initial compositions. We therefore come to the conclusion that the creation of small compartments can allow the formation of a population of different individuals out of the fluctuations of an environment that looks macroscopically homogeneous. While not yet sufficient, this is definitely a necessary condition for darwinian evolution to take place (obviously supposing that the compartments can divide and that their division rate depends upon composition).
Moreover, in small stochastic systems it may also happen that there are different trajectories stemming from the same initial conditions, due to the order in which new molecules are synthesized.

These aspects of protocell dynamics are very important and our model, in spite of its current limitations, is well suited to explore the related phenomena. It is indeed possible to analyze the different possible stochastic effects, that include:

\begin{enumerate}[(i)]
\item the path dependency induced by the random order in which new molecules are generated, with particular regard to the low concentration effects: if a catalyst is discovered in advance with respect to another, the system evolution may be different; this can be studied by comparing different simulations referring to the same chemistry, starting from the same initial conditions;
\item the differences induced by different initial conditions, randomly generated from the same distribution;
\item the different behaviors of distinct chemistries: in the real world the rules of the chemistry are given, but in the kind of analysis performed here it is also interesting to understand how different chemistries may affect the behavior of the system and the diversity of a population of protocells. As, for example, the role of {\em RAF sets} \citep{Hordijk2010,Hordijk:2004vl,Hordijk2012} in the overall dynamics.
\end{enumerate} 
 
In Section~\ref{sec:results} we present the results of the simulations on the model.
In the concluding Section~\ref{sec:final} we will discuss the main findings of this paper, and we will propose further analysis and refinement of the models.

\newpage
\section{Main features of the model}
\label{sec:model}
\paragraph{Entities and interactions.}
\label{sec:entInt}
The model describes an open system in which simple molecules interact with each other through elementary catalyzed reactions. The basic entities of the system are monomers and polymers, identified by ordered strings of letters oriented from left to right taken from a finite alphabet (e.g. $A, B, C, \ldots$). We will refer to the\lqu letters\rqu~also as bricks, while the term monomer will be reserved to those molecular types that are composed of a single brick. 

Every species $x_i,i=1,2,\ldots,N$ composing the entire set of species $X=[x_1,x_2,\ldots,x_N]$ is characterized by its specific amount (either quantity or concentration), denoted by $\hat{x}_{i}$. The number of copies of a specific species defines its number of molecules. 

The two basic reactions are: \textit{i)} cleavage, i.e. the cutting of a species composed of more than one brick into two shorter species (e.g. $ABBA \rightarrow A + BBA$) and \textit{ii)} condensation, i.e. the concatenation of two species in a longer one (e.g. $BBA + AAB \rightarrow BBAAAB$).

A key assumption of the model is that any reaction occurs only if catalyzed by one of its specific catalysts; hence we exclude the presence of spontaneous reactions by assuming sufficiently high activation energy for each reaction. In particular, condensation requires an intermediate reaction in which a temporary complex between the catalyst and the substrate is formed.

A further restriction regarding catalysis imposes that only species that are composed of at least a minimum number of bricks can be catalysts. Additionally, we neglect the presence of backward reactions (exception made for the dissociation reaction of intermediate complexes) by hypothesizing that the Gibbs energy $\Delta G$ for any reaction is sufficiently large.
A schematic representation of the various reaction types is given in Fig. 1.\\ 
\begin{figure}[htp]
\centering
	\begin{tabular}{c}
		\includegraphics[scale=0.3]{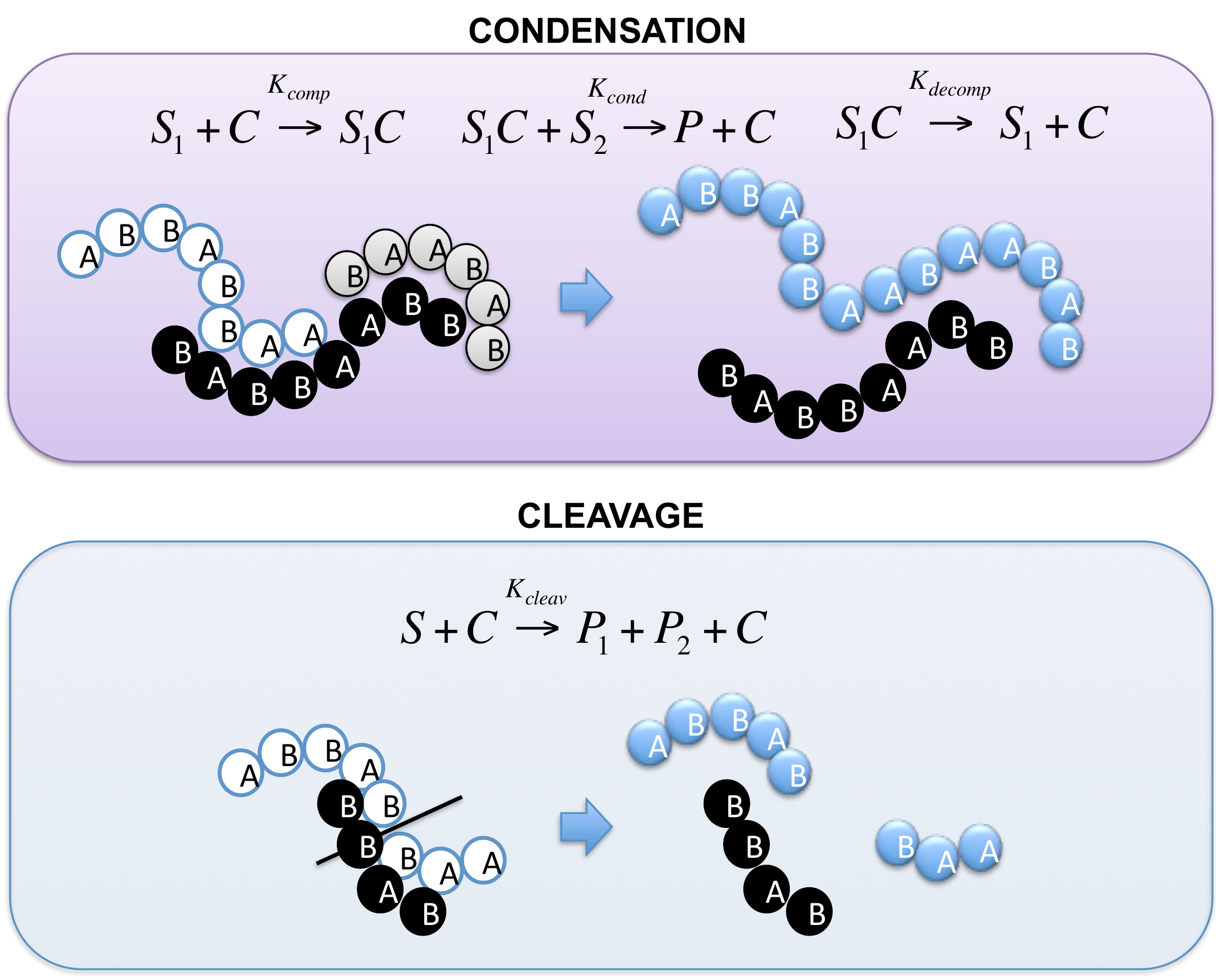}
	\end{tabular}
	\label{fig:rctscheme}
	\caption{Schematic representation of the two possible reaction types, namely condensations and cleavages. $S, S_1$  and $S_2$ represent the substrates of a reaction, $C$ is a species with catalytic function, $S_{1}C$ is the transient complex that is formed in the first step of the condensation and $P, P_1$ and $P_2$ stand for the products of a reaction. $K_{cleav}, K_{comp}, K_{diss}$ and $K_{cond}$ respectively stand for the kinetic rates of cleavage, complex formation, complex dissociation and final condensation.}
\end{figure}
\paragraph{The reaction network.}
\label{sec:rctnet}
Provided that a catalyst for the reaction exists, each species in the system could condensate with any other species in the system, or be split at any cutting point into smaller species. The cardinality of the set $R$ of all the conceivable reactions for the system, at a given time, is therefore given by:

\begin{equation}
\hat{R}=\sum_{i=1}^{N}(L_{(x_{i})}-1)+N^{2}
\label{eq:conceivableRcts}
\end{equation}

where $N$ is the cardinality of $X$ and $L_{(x_{i} )}$ is the length of $x_i$ (i.e. the number of bricks of that specific species). The first term of Eq.~\ref{eq:conceivableRcts} refers to the conceivable cleavages and the second one to the conceivable condensations. Hence, the set of reactions that are actually allowed, i.e. the~\lqu chemistry\rqu~of the system, is determined by the catalysts involved in the various reactions.
Following Kauffman, we define the chemistry of the system at random. In this regard, any species $x_i$ has a finite fixed probability $p_i$ of being chosen as catalyst of a randomly chosen reaction among those belonging to $\hat{R}$. It is worth stressing that, although the reaction network is built probabilistically, it remains invariant in time; in other words, once a species is chosen to be the catalyst of a given reaction, that species will always be catalyst for that reaction. Further details concerning the structures of the model can be found in~\citep{Filisetti:2010fk,Filisetti2011b}.

\paragraph{The dynamics.}
\label{sec:dynamics}
The system's dynamics is simulated by means of an extension of the well-known Gillespie algorithm~\citep{Gillespie:2007vn,Gillespie:1977fv} for the stochastic simulation of chemical reaction systems, in which we allow the appearance of novel species and reactions that are not present in the system in the initial conditions. In fact, cleavage and condensation of elements either present within the protocell or entering it from the external environment can generate new species, which, in turn, can be involved in new reactions, both as catalysts and substrates.
	
\begin{figure}[htp]
\centering
	\begin{tabular}{c}
		\includegraphics[scale=0.4]{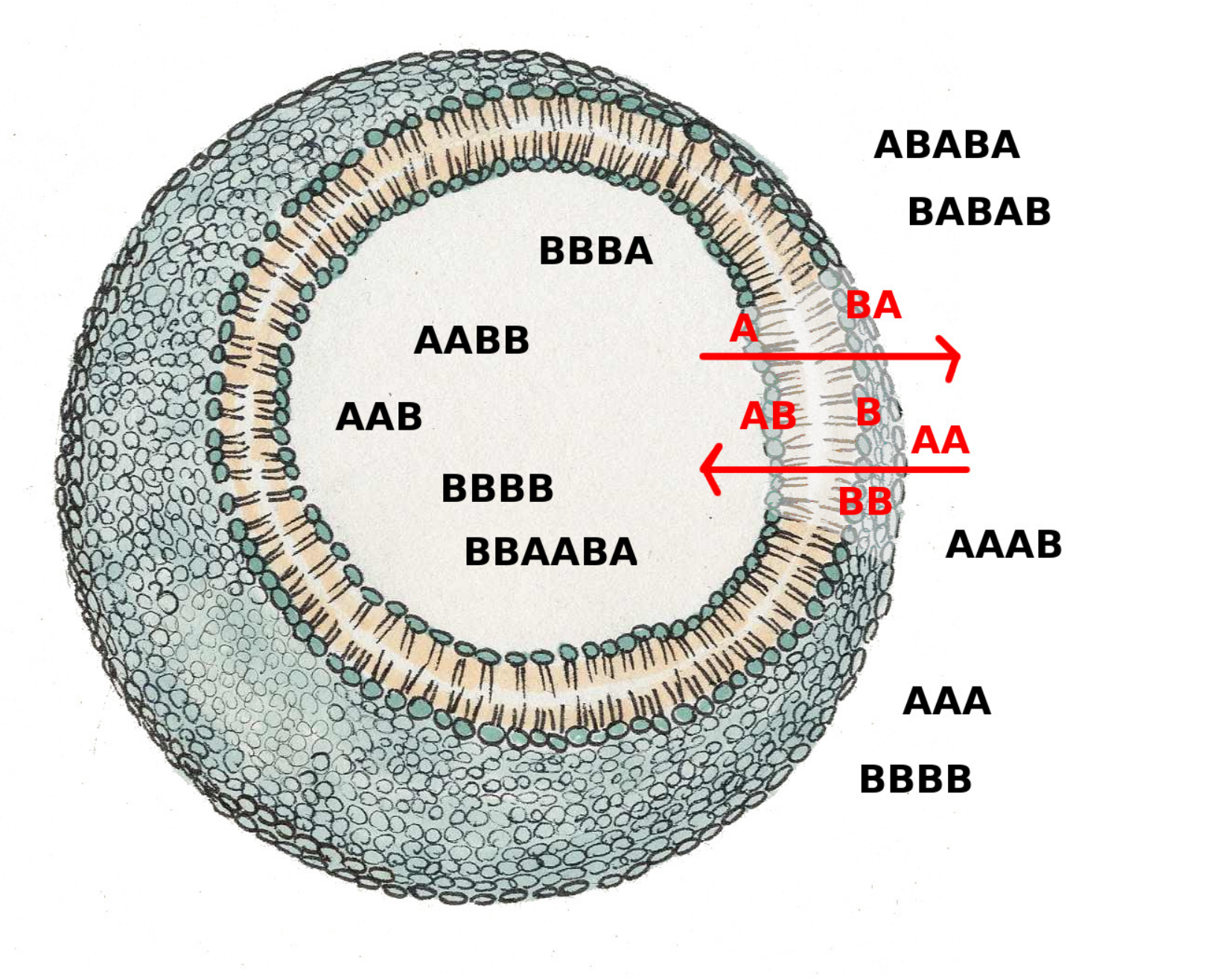}
	\end{tabular}
	\label{fig:protocell}
	\caption{Schematic representation of the semi-permeable membrane as conceived in our model of protocell. The membrane is here represented as a lipid bilayer that shapes a spherical vesicle. In this example, only the species shorter than 3 letters can cross the membrane, entering or leaving the internal compartment, the dynamics of which is actually simulated in our model of catalytic reaction networks. On the opposite, longer species are confined either inside or outside the vesicle.}
\end{figure}

\paragraph{The container: the introduction of the protocell.}
\label{sec:protocell}
In its classical formulation~\citep{R.J.Bagley:1991ys}, the catalytic reaction network is modeled within a continuous stirred-tank reactor (CSTR), in which the ingoing and outgoing fluxes could be adjusted according to the experimental needs. In this work, we introduce a major modification of the model by introducing a semi-permeable membrane that separates the catalytic reaction network from the external environment (see Fig.~\ref{fig:protocell}).\\
The semi-permeable membrane is here modeled by allowing only some species to enter and leave the protocell, namely those that are shorter than an arbitrary length $L_{perm}$. All the species that are longer than $L_{perm}$ either remain entrapped within the protocell or never enter it from the outside. Another important assumption is that the concentration of the permeable molecules is homogeneous both inside and outside the protocell, and that they take the same value. That is, we assume infinitely fast diffusion both in the bulk phases and across the membrane, so that the chemical potentials of the permeable species are the same. The species that can cross the membrane will be also defined, from now on, as the {\em buffered} species. 

In this work, we consider the volume of the protocell to be constant, while it is planned to introduce protocell size and duplication dynamics in further developments of the model. 

We also remark that in this model the transport processes in a vesicle are treated in a way that, albeit simplified, parallels their actual dynamics. While small flow reactors have also been proposed as protocell models~\citep{Hordijk2012,Vasas2012}, they are indeed not well suited, since they require a constant inflow that has no physical analogue in a vesicle, and usually also allow the outflow of all the solutes. 

\begin{algorithm}[H]
\caption{Formal definition of the model as multiset-rewriting system.}
\label{alg:driver}
\begin{algorithmic}[1]
\STATE {let $\Psi$ be the alphabet of symbols denoting monomers, $A, B, C ...$;}
\STATE {the  $\Theta^*$ and $\Xi^*$  be the (infinite) set of  all  finite-length non-empty {\em polymers} and  {\em substrate-catalyst  complexes} over $\Psi$, and
 let us denote by $P \cdot Q$ (resp. $P \_Q$, substrate-catalyst) the polymer (resp. complex) obtained by concatenating polymers $P$ and $Q$;}
\STATE{ let the state of the system be  multiset $\Phi\equiv  P_1 \mid P_2\mid...\mid P_n$ counting the occurrences of each polymer/complex $P_i \in \{\Theta^* \cup \Xi^*\}$;}
\STATE consider the following reaction schema with variables  $X, Y$ and  $Z$, in a multiset-rewriting notation:
\\
\begin{itemize}   \itemsep1pt \parskip0pt \parsep0pt
\item {\bf [A: Cleavage]} \\ $R_{cl}: X \cdot Y \mid Z\mapsto X \mid Y \mid Z$ \\
\COMMENT{$Z$ is the catalyst, $X \cdot Y$ is the polymer breaking to $X$ and $Y$}
\\
\item {\bf [B1: Complex formation]} \\$R_{cf}: X|Y \mapsto X \_ Y$ \\
\item {\bf [B2: Complex dissociation]} \\ $R_{cd}: X \_ Y \mapsto X\mid Y$
\item {\bf [B3: Final condensation]} \\ $R_{fc}: X \_ Y \mid Z \mapsto X \cdot Z \mid Y$\\
\COMMENT{$X$ and $Z$ are substrates, $Y$ is the catalyst, $X\_Y$ is the substrate-catalyst complex, $X \cdot Z$ the polymer composed by $X$ and $Z$}

\end{itemize}
Variable polymers ($X \cdot Y$)  take values in $\Theta^*$, whereas complexes  ($X \_Y$)   in $\Xi^*$.

\STATE (Reactions generation)  an {\em instantiation map} $\varphi$ is evaluated via standard pattern matching to transform the above schema in all the possible rewriting rules 

\[
{\cal R}^{\varphi} = R_{cl}^{\Phi}\cup R_{cf}^{\Phi} \cup R_{cd}^{\Phi} \cup R_{fc}^{\Phi}
\]

for multiset $\Phi$. The map $\varphi$, which assigns values to $X, Y$ and  $Z$, is required to be {\em maximal} and {\em univocal}, i.e. it must yield all the possible rewriting rules applicable in $\Phi$  and instantiated consistently (via deterministic association) for any of its evaluation.  
\STATE  $\Phi$ and ${\cal R}^{\varphi}$ yield a Continuous-time Markov Chain.
By firing a reaction in ${\cal R}^{\varphi}$ via the classical Gillespie algorithm a new state of the system is generated. In this particular implementation of the model, polymers shorter than 3 are fed and cannot be instantiated as catalysts (see Appendix \ref{app:app1}). The simulation proceeds with the new state from previous step. 
\end{algorithmic}
\end{algorithm}

\section{Results of the simulations}
\label{sec:results}
The preliminary experiments on the protocell model were performed by keeping fixed some key structural parameters (see the Appendix for the details) and creating different random chemistries. 
In particular, we here present the details of two specific chemistries, which were specifically selected among many (on a total of 10 simulated chemistries) because they exhibit very specific dynamical properties.
In this regard, it is first important to introduce the concept of {\em RAF set}, which be fundamental in the description of these systems. 

 Following Hordijk et al.~\citep{Hordijk2010,Hordijk:2004vl,Hordijk2012}, given the entire chemistry, a subset of reactions $\mathcal{R}$ is: $(i)$ \textit{Reflexively autocatalytic} (RA) if every reaction in $\mathcal{R}$ is catalyzed by at least one other reaction belonging to $\mathcal{R}$, $(ii)$
 \textit{Food-generated} (F) if every substrate in $\mathcal{R}$ can be produced from the food set \textit{F} by means of the reactions belonging to $\mathcal{R}$, $(iii)$ \textit{Reflexively autocatalytic and F-generated} (RAF) if both the previous conditions are satisfied.

The chemistries of the two presented protocells differ for the presence/absence of a RAF set (RAF in short from now on): in particular, in the first protocell ($CH1$) no RAFs are present, whereas in the second protocell ($CH2$) a RAF formed by an autocatalysis consuming molecules from the food set is present.

In order to assess the behavior of the system, a method to measure the {\em similarity} between two different states of the system is that proposed by Kaneko~\citep{Kaneko:2006eu} and is based on the comparison of the vectors that describe their chemical compositions. 
Let us define the k-dimensional vectors $C_{j}(t)=[c_{j,1}(t),c_{j,2}(t),...,c_{j,N} (t)]$ and $C_{k}=[c_{k,1} (t),c_{k,2} (t),...,c_{k,N} (t)]$ whose components are the concentrations of the species present at time $t$ in systems $j$ and $k$ respectively. Same positions in vectors refer to same species, hence species present in system $j$ that are not present in $k$ have concentration equal to 0 and viceversa. 
The similarity between the two systems is then computed by means of the normalized inner product:
%
%
\begin{equation}
\Theta_{t}=cos^{-1}\left(\frac{\vec{C_j}(t) \cdot \vec{C_k}(t)}{||C_j(t)|| \cdot ||C_k(t)||}\right)
\label{eq:angle}
\end{equation}

where $\Theta_{t}$ is the angle between the two vectors measured at the time $t$. Throughout this paper, angles will be measured in degrees.

\subsection{Low concentration effects}
\label{sec:pathDep}

In this section we investigate the low concentration effects in protocells. To this end, the selected chemistries are tested with respect to 4 different initial uniform concentrations of the chemical species present inside the vesicle: in detail, the initial concentration of all the non-buffered species is equal to: $1 mM$, $0.1 mM$, $0.01 mM$ and $1 \mu M$. Conversely, the amount of each buffered species is always fixed to $1 mM$.

Notice that, in general, the protocell model presented here does not lead to stationary chemical distributions. Indeed, situations are possible where some particular species indefinitely increases its concentration: an example is the RAF set in $CH2$, where a particular species catalyses its own formation using two buffered species as substrates: in this case an exponential growth is achieved. Another simple case could be the linear growth of a species produced directly from the buffered species by means of a catalyst not involved in other reactions, i.e. present in constant concentration. Of course, more complex situations are possible, even not directly originating from the food set.
Yet, many simulations show a transient in which the chemicals distribution rapidly change followed by a long quasi-equilibrium, where changes are limited to small adjustments. By applying the parameters adopted in the present work, such a state is already achieved in $2500$ seconds, except for the case of very low concentrations ($1\mu M$).
\begin{center}
\begin{table}
\footnotesize
  \begin{tabularx}{\textwidth}{|c|X|X|X|X|X|X|X|}
    \hline
    Conc & Mol- ecules per species &
    \multicolumn{2}{ |X| }{$CH1$} &
    \multicolumn{2}{ |X| }{$CH2$} &
    \multicolumn{2}{ |X| }{$CH2$ (no RAF)}\\
    \hline
    Molarity & Average & $\Theta_{3000}$ (mean) & $\Theta_{3000}$ (max.) & $\Theta_{3000}$ (mean) & $\Theta_{3000}$ (max.) & $\Theta_{3000}$ (mean) & $\Theta_{3000}$ (max.) \\ 
    \hline
    (Cond.1) $1mM$ & 600 & 0.41 & 0.68 & 0.06 & 0.19 & 0.57 & 0.96   \\ 
    \hline
    (Cond.2) $0.1mM$ & 60 & 2.34 & 5.86& 0.18 & 0.52  & 1.69 & 2.78  \\ 
    \hline
    (Cond.3) $0.01mM$ & 6 & 7.71 & 15.28 & 9.69 & 21.86 & 6.48 & 11.21  \\ 
    \hline
    (Cond.4)
 $1\mu M$ & 1 & 11.15 & 19.35 & 3.67 & 11.91 & 9.44 & 15.35  \\ 
    \hline
\end{tabularx}
 \caption{In the table the average and the maximum values of $\Theta_{3000}$ relative to 10 distinct simulations of each of the $4$ different initial conditions (rows in the table) are shown. The measures are reported for the  2 different chemistries: the one without RAFs and the one with RAFs, for which they are also computed without taking into account the species belonging to the RAF, column~\lqu CH2 (no RAFs)\rqu. The $4$ conditions differ in the average magnitude of the initial concentrations of the initial set of molecular species not belonging to the buffered flux (the food set). A realization of each of the 4 initial concentration is drawn at random from a Poisson distribution, according to the given parameters, and is maintained invariant across the 10 different runs.}
   \label{tab:angleALL}
 \end{table} 
  \end{center}

\paragraph{Diversity.}In table~\ref{tab:angleALL} some statistics on how $\Theta$ varies according to the different cases are shown and, in particular, it is possible to appreciate an increase of the dissimilarity among protocells in correspondence with a decrease on the initial concentration, as indicated by the average and the maximum value of $\Theta$ between all the runs of each specific chemistry and specific initial condition, computed at time $t=3000$.\footnote{Exceptionally for the case $1\mu M$,  $\Theta$ is not computed after 3000 second but when at least $5000$ reactions have occurred within the simulation. The reason for this is that the low concentrations involve a so slow dynamics that $3000$ seconds are not enough in order to observe significant chemical changes.} Since these runs are characterized by identical chemistries and identical initial conditions, the angle reported here is indeed the result of dynamical evolutions that differ only for the simulation random seed. It is worth remarking that the lower is the concentration profile, the higher is the distance among the final distributions,\footnote{In regard to $CH2$ we are here considering the value of $\Theta$ excluding the species belonging to the RAF set (last columns of the table). Since the molecules belonging to the RAF set reach a concentration much greater with respect to the other molecules, considering them in the angle computation would misrepresent the distance among the simulations.} i.e. the angle $\Theta$ reaches its maximum value with regard to very low initial concentrations (e.g. $1$ molecule per species on average).

It is interesting to notice that the displayed low concentration effects with respect to the overall similarity of the system are found in both chemistries, hinting at a  generic property of such systems, independently of the possible presence of RAF sets.

\begin{figure}[htp]
\centering
	\begin{tabular}{c}
		\includegraphics[scale=4.8]{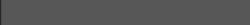}\\
		\includegraphics[scale=4.8]{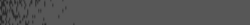}\\
		\includegraphics[scale=4.8]{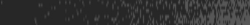}\\
		\includegraphics[scale=4.8]{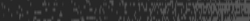}\\
	\end{tabular}
	\caption{From the top to the bottom the four traces, accounting for conditions $1, 2, 3$ and $4$ of the chemistry 2, i.e. the one containing the RAF set, depicting the presence of the RAF during the simulation (time flows from left to right, each row represents a different run of the same condition). The black zones stands for the absence of the RAF set, while gray zones denote the presence of the RAF. It is important to remark that RAFs are searched in a dynamic network created with the reactions occurred in a mobile time window of 5 seconds. }
	\label{fig:traces}
\end{figure}

\paragraph{The influence on RAF set dynamics.} The asynchronous framework implies that only one reaction occurs at the time. Given that RAFs in general involve more than one reaction, in order to detect their presence we analyze the system in a time interval \textit{(i)} sufficiently large to let a significant number of reactions to occur, yet \textit{(ii)} not embracing the whole simulation (in order to avoid the presence of too rare reactions); It is worth stressing that since the analysis are made {\em ex post} with respect to the simulation of the system, they do not affect the simulated dynamics, but just the way of representing it. In fig.~\ref{fig:traces}, we show the presence (or the absence) of the RAF of $CH2$ every $10$ seconds, by using a time window of $5$ seconds,\footnote{We set the sampling frequency and the time threshold of the windows by taking advantage from several initial model threads, not essential to the comprehension of this article.} with respect to the 4 aforementioned different average concentrations.\\
Note that in all the initial conditions the autocatalytic species has a concentration higher than zero, thus the corresponding RAF is always able to achieve its own growth viability.
From Fig.~\ref{fig:traces} it is possible to observe the strong correlation between the average concentration of the species within the protocell and the presence of RAFs. As long as the concentration decreases, so does the probability of detecting a RAF, yet in the long run in most of the simulations RAFs eventually emerge. 
 
\subsection{Sensitivity to initial conditions}
\label{sec:difInitConc}

We here analyze the effects of variations in the initial concentrations of the single species present within the protocells, while maintaining their average concentration fixed. We simulate 10 different variations of the single species concentrations for both the case of average concentration equal to $0.01mM$ (condition $3$) and average concentration $1\mu M$ (condition 4).

\begin{figure}[htp]
\centering
	\begin{tabular}{c}
		\includegraphics[scale=0.3]{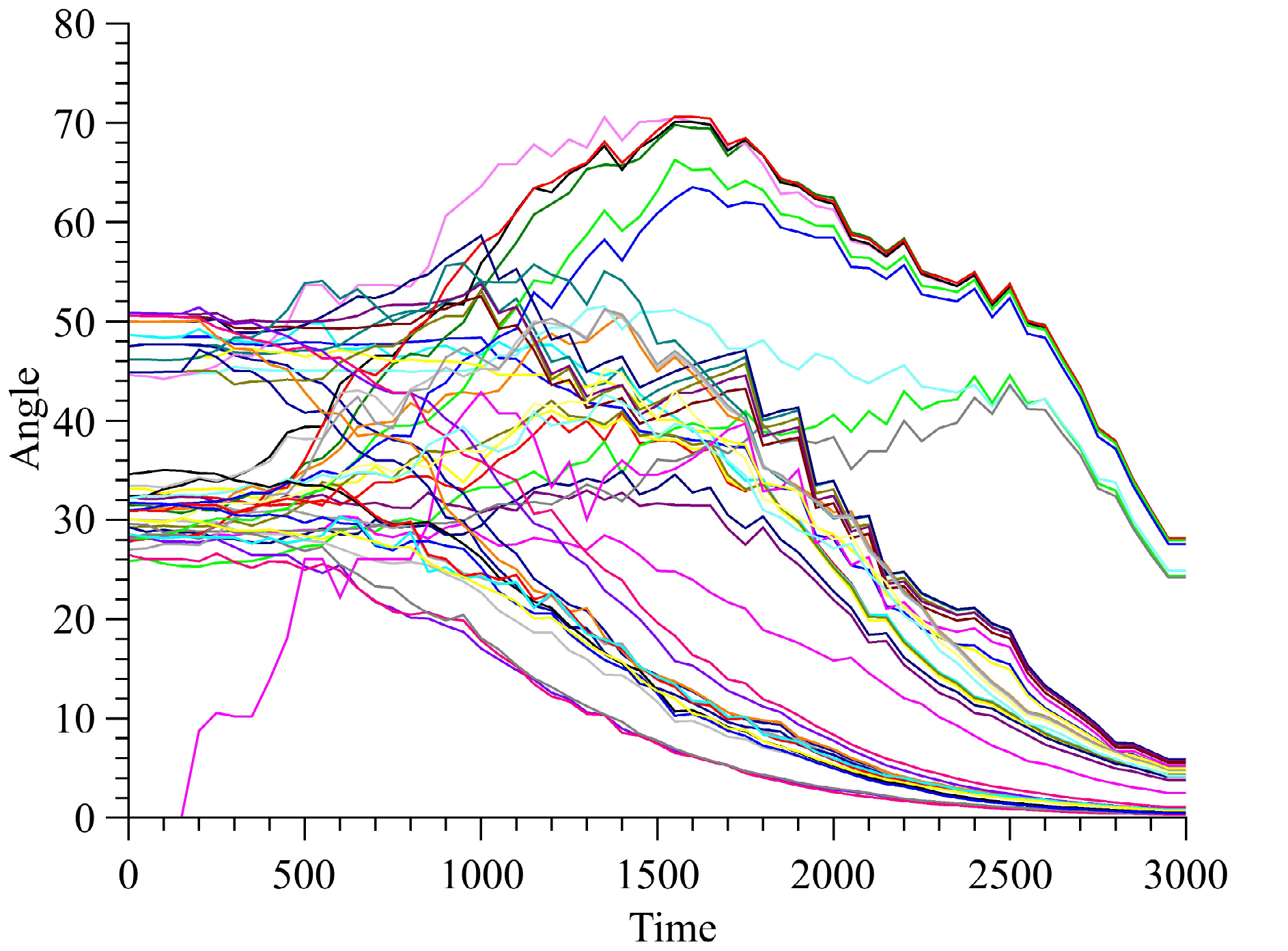}{\tiny(a)}
		\includegraphics[scale=0.3]{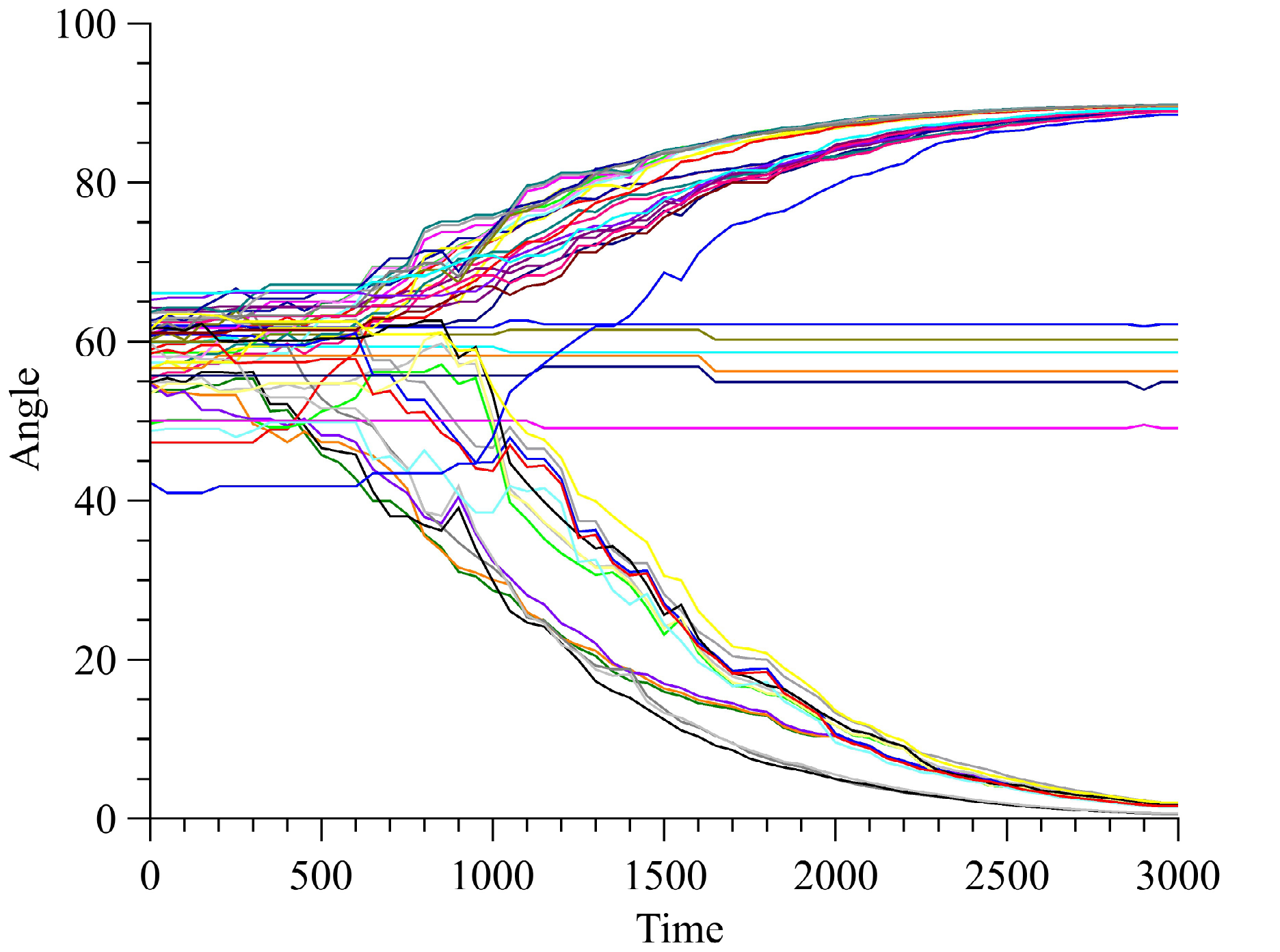}{\tiny(b)}\\
	\end{tabular}
	\caption{The angles between each couple of different simulations in time, for condition $3$ ($0.01mM$ - left panel) and condition $4$ ($1\mu M$ - right panel).}
	\label{fig:angleintime}
\end{figure}

In Fig.~\ref{fig:angleintime} we display the variation of the similarity in time for each couple of simulations, with respect to both conditions, in order to provide a picture of the sensitivity of the overall diversity to the initial conditions. 

Note that the very low concentrations of condition $4$ (in average, only one molecule for each species) imply that in certain protocell realizations some species are missing: in particular, each simulation starts from a different set of species composed of, on average, $40$ species, on the possible $62$. This may explain the very high values of $\Theta_{0}$ with respect to condition $4$ (Fig. \ref{fig:angleintime}).

Besides, one can see that condition $4$ shows an evident bifurcation (Fig.~\ref{fig:angleintime}b). Given that in some simulations, because of the low concentration effects, the autocatalytic species (and so the RAF) can be present or absent in the initial condition, the system can indeed reach different regions of the state space, leading to deeply different histories. On the opposite, condition $3$ shows the aforementioned regulatory activity of the always active RAF. 

\begin{figure}[htp]
\centering
	\begin{tabular}{c}
		\includegraphics[scale=0.28]{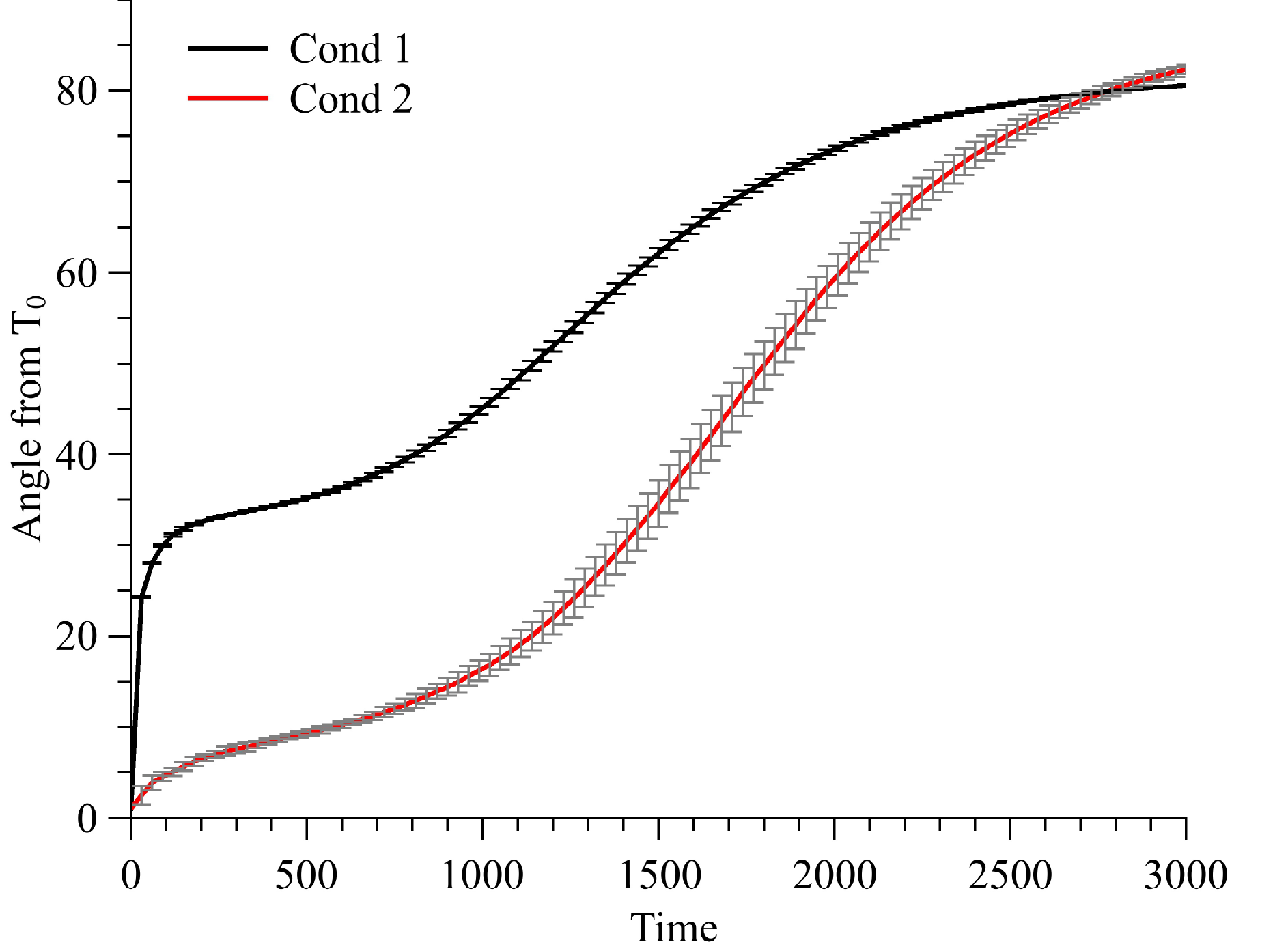}{\tiny(a)}
		\includegraphics[scale=0.28]{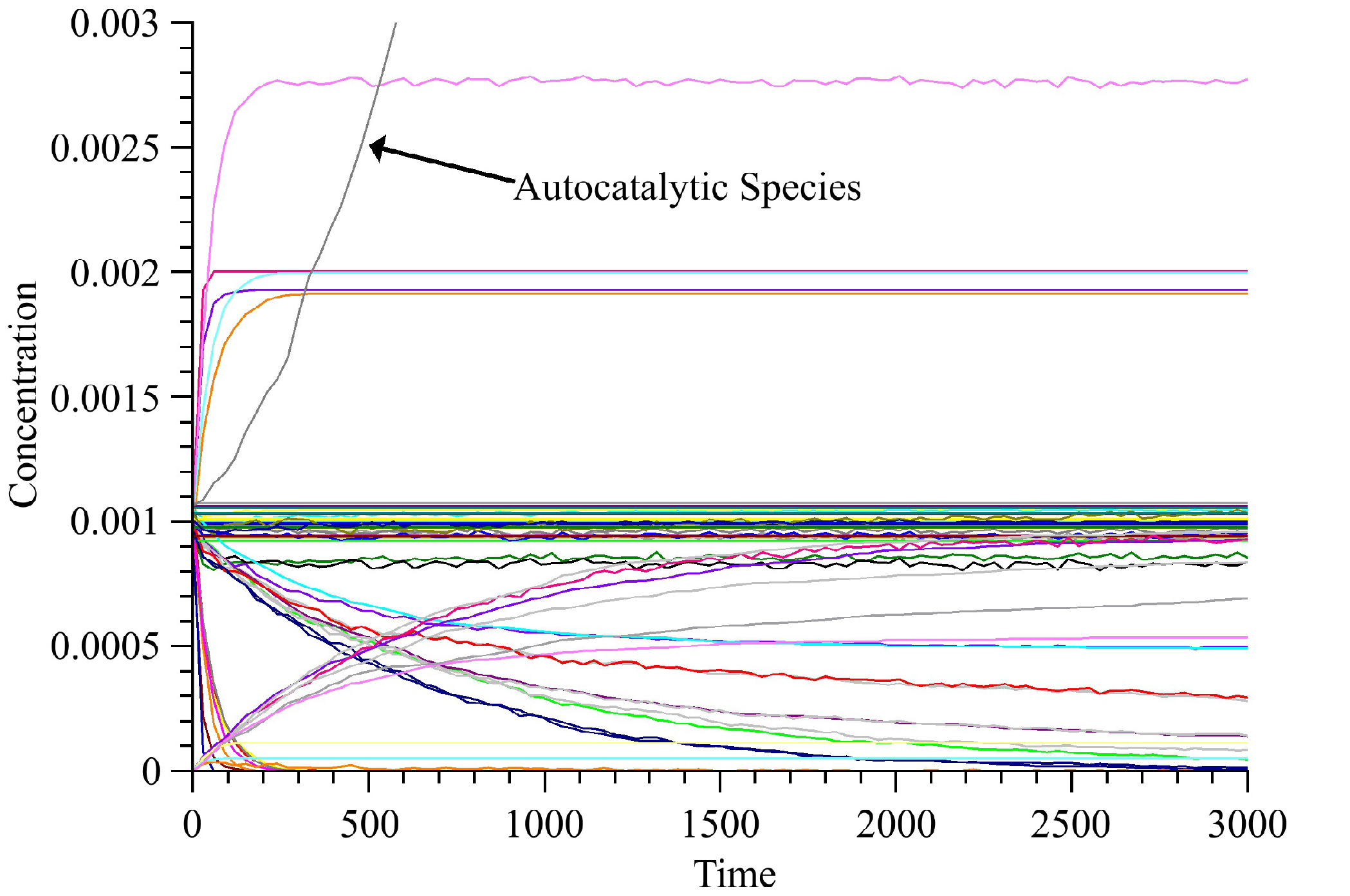}{\tiny(b)}\\
	\end{tabular}
	\caption{On the left panel the average angle (and the standard deviation) $\Theta_{T_{0}\rightarrow T_{t}}$ measured from $T_{0}$ to $T_{t}$ for each chemical distribution, within each single run of conditions $1$~(average molarity $1mM$) and $2$~(average molarity $0.1mM$), is shown. It is worthwhile to remark that $\Theta=90$ stands for a complete orthogonality between the chemical distributions. In order to appreciate the convergence towards a quasi-orthogonality, on the right panel the concentration of the species in time of a particular simulation are shown (Simulation: CH2, condition 1, run 1). The exponential growth concentration of the autocatalytic species is shown with a limit of $3mM$.}
	\label{fig:anglefromt0}
\end{figure}

Finally, it is important to remark that during each simulation many species disappear and many other appear, so a compact way to follow the process is that of monitoring, during each run, the variation of the angle $\Theta_{T^{0}\rightarrow T^{t}}$ between the initial and the current chemical setting (Fig.~\ref{fig:anglefromt0}a). The graph indicates that $\Theta_{0 \rightarrow 3000}$ approaches a value close to $80$ degrees, which means a quasi-orthogonality of the system with respect to the initial condition. 

The dynamics of the system, in term of molecular concentrations, leading to such a divergence from the beginning is shown in Fig.~\ref{fig:anglefromt0}b. It is reasonable to hypothesize that the clear inversions in the trajectories of the concentrations of certain species may be responsible for this phenomenon. 

\section{Conclusion and further developments}
\label{sec:final}
In this paper we  introduced a simplified model of a non-growing protocell and we  investigated the behavior of a stochastic model of catalytic reaction networks in such an environment. To the best of our knowledge this is a novel approach.

The crucial importance of the small size of the protocell has been stressed, and the effects of the fact that some chemicals can be present in low numbers have been investigated. While a broader analysis is ongoing, we have here shown that it is possible to reach different compositions of the chemical species, in the particular case in which some species are present in the bulk at low concentrations. We have also shown that there are two different, possibly overlapping reasons for this diversity: $(i)$ the random sequence of molecular events involving those species and $(ii)$ the random differences in their initial concentrations. We have also stressed the importance of RAF sets in influencing the overall dynamics.

There are several ways in which this work might seed further research.
The most obvious is that of relaxing the physical limitations that have been considered, e.g. infinitely fast diffusion, yet we do not except that this may change the major conclusions summarized above.

Obviously, a very interesting direction is that of considering a protocell that is able to grow and divide. The processes involved in protocell growth and replication are indeed complex and, in particular, a necessary condition for its existence and replication is the coupling between the rates of molecules replication and cell growth. We have shown elsewhere~\citep{Carletti:2008rm,Filisetti2010b,Serra:2006aa} that the very existence of this coupling suffices to guarantee (under very general conditions) that, in the long run, the rate of cell division and that of duplication of the replicating molecules converge to the same value, thereby allowing sustainable growth of a population of protocells.\footnote{This property was proved earlier by Munteanu et al. for the Los Alamos bug mode~\citep{Munteanu:2006aa}.} However, those results were achieved by supposing a fixed set of genetic memory molecules, with some possible extinction. It could be sound to extend this approach to the case where there are evolving chemical reaction sets and to verify whether synchronization occurs.
An important aspect to be addressed in the case of growing vesicles is also the effect of volume growth on the concentrations of the various chemicals: a preliminary investigation can be found in~\cite{Carletti2012}. 

Besides, we have not here explicitly considered the possibly catalytic role of membranes that, as it has been discussed in section~\ref{sec:intro}, might be a major cause of the difference between the intracellular environment(s) and the bulk. In a fixed volume model this effect can be lumped in the effective reaction rates, but if we consider a growing protocell we have to take into account the differences between surfaces and volumes. This might also lead to interesting phenomena that will be analyzed in future developments.

To conclude, different protocells may host different mixtures of molecular species, even if they share the same chemistry (i.e., they~\lqu inhabit the same world\rqu). It might be extremely interesting to model the behavior of populations of different protocells of this kind, which may show different growth rates, but may also undergo phenomena like coalescence, exchange of material, etc.~Thus, further investigations will be indeed necessary to assess different generations of protocell populations and their possible evolution pathways.\\
Last but not least, it will be interesting to extend these studies to other protocell architectures like e.g. the Los Alamos bug, Gard models, and others.

\section{Acknowledgements}
Stuart Kauffman, Norman Packard and Wim Hordijk kindly shared with us their deep understanding of autocatalytic sets in several useful discussions. Useful discussions with Ruedi F\"uchslin, Davide De Lucrezia, Timoteo Carletti, Andrea Roli and Giulio Caravagna are also gratefully acknowledged. The authors are also grateful to Giulia Begal for kindly drawing the image of Fig.~\ref{fig:protocell}. C.D. wishes to acknowledge the project SysBionet (12-4-5148000-15; Imp. 611/12; CUP: H41J12000060001; U.A. 53) for the financial support of the work.\\
The final publication is available at Springer via http://dx.doi.org/10.1007/s11047-014-9445-6\\\\

\appendix
\section{Appendix 1 - Simulation environment and parameter settings}
\label{app:app1}
Simulations were performed with the CaRNeSS simulator\footnote{https://github.com/paxelito/carness} developed by the research group.\\ 
In the following, the baseline setting of the system used in the simulations is reported (for the parameters that were variated in the different experiments please refer to the text):
$(\bullet)$ Alphabet: A, B, $(\bullet)$ Volume = $ = 1e-18 dm^3 = 1\mu^3$,  $(\bullet)$ Average catalysis probability = 1 catalyzed reaction for species,  $(\bullet)$ Maximum length of the species, $L_{max} = 6$,  $(\bullet)$ $L_{perm} = 2$,  $(\bullet)$ Monomers and dimers do NOT catalyze,  $(\bullet)$ $K_{cleav}= 25M^{-1}sec^{-1}$, $(\bullet)$  $K_{comp}=50M^{-1}sec^{-1}$, $(\bullet)$ $K_{diss}=1M^{-1}sec^{-1}$, $(\bullet)$ $K_{cond}=50M^{-1}sec^{-1}$.

\footnotesize
\bibliographystyle{plain}

\end{document}